\documentclass[usenatbib,onecolumn]{mnras}
\usepackage{graphicx}
\graphicspath{{./}{./figures/}}

\newcommand*{\mysub}[2]{\ensuremath{#1_{\mathrm{#2}}}}

\newcommand*{\expectation}[1]{\ensuremath{\left\langle #1 \right\rangle}}
\newcommand*{\ltsim}{\ {\raise-.75ex\hbox{$\buildrel<\over\sim$}}\ }
\newcommand*{\gtsim}{\ {\raise-.75ex\hbox{$\buildrel>\over\sim$}}\ }
\newcommand*{\proptosim}{\ {\raise-.75ex\hbox{$\buildrel\propto\over\sim$}}\ }
\newcommand*{\normal}{\ensuremath{\mathcal{N}}}

\newcommand*{\like}{\ensuremath{\mathcal{L}}}
\newcommand*{\Lsimp}{\mysub{\like}{sim}}
\newcommand*{\Ldet}{\mysub{\like}{det}}
\newcommand*{\ylim}{\mysub{y}{1,lim}}

\newcommand*{\xlim}{\mysub{x}{lim}}
\newcommand*{\Pdet}{\mysub{P}{det}}
\newcommand*{\Ndet}{\mysub{\hat{N}}{det}}
\newcommand*{\Nmis}{\mysub{N}{mis}}
\newcommand*{\fdet}{\expectation{\mysub{f}{det}}}
\newcommand*{\fmis}{\expectation{\mysub{f}{mis}}}
\newcommand*{\EN}{\expectation{N}}
\newcommand*{\ENdet}{\expectation{\mysub{N}{det}}}
\newcommand*{\ENmis}{\expectation{\Nmis}}

\newcommand*{\lrgs}{{\sc lrgs}}
\newcommand*{\trunc}{{\sc lrgs.trunc}}

\begin{document}

\title[A Primer on Regression with Truncated Data]{Coping with Selection Effects:\\ A Primer on Regression with Truncated Data}

\author[A.\ B.\ Mantz]{Adam B.\ Mantz$^{1,2}$\thanks{Corresponding author e-mail: \href{mailto:amantz@stanford.edu}{\tt amantz@stanford.edu}}\\
  $^1$Kavli Institute for Particle Astrophysics and Cosmology, Stanford University, 452 Lomita Mall, Stanford, CA 94305, USA\\
  $^2$Department of Physics, Stanford University, 382 Via Pueblo Mall, Stanford, CA 94305, USA
}
\date{Submitted 24 July 2018, accepted 28 January 2019}

\maketitle

\begin{abstract}
  The finite sensitivity of instruments or detection methods means that data sets in many areas of astronomy, for example cosmological or exoplanet surveys, are necessarily systematically incomplete. Such data sets, where the population being investigated is of unknown size and only partially represented in the data, are called ``truncated'' in the statistical literature. Truncation can be accounted for through a relatively straightforward modification to the model being fitted in many circumstances, provided that the model can be extended to describe the population of undetected sources. Here I examine the problem of regression using truncated data in general terms, and use a simple example to show the impact of selecting a subset of potential data on the dependent variable, on the independent variable, and on a second dependent variable that is correlated with the variable of interest.
Special circumstances in which selection effects are ignorable are noted.  
I also comment on computational strategies for performing regression with truncated data, as an extension of methods that have become popular for the non-truncated case, and provide some general recommendations.
\end{abstract}


\section{Introduction}

As astronomers increasingly adopt Bayesian methods, and computational resources continue to improve, ubiquitous features of astronomical data and models that are difficult to address through classical statistical methods based on the generalized linear model are now routinely dealt with. Among these are measurement errors on the independent variables of a regression, correlation in the measurements of independent and dependent variables (hereafter called covariates and responses, respectively), and the presence of intrinsic scatter. In addition to general-purpose Markov Chain Monte Carlo tools, easy-to-use codes for specialized but reasonably generic problems have been provided to and socialized within the community. Of particular note is {\sc linmix\_err} \citep{Kelly0705.2774}, which uses conjugate Gibbs sampling to efficiently fit a model consisting of a linear mean relation, Gaussian measurement and intrinsic scatters, and a Gaussian mixture prior distribution of the covariates. While these are strong modeling assumptions, many of them are also fairly common, irrespective of the fitting method employed. The same approach has been generalized to multivariate regression (\lrgs{}; \citealt{Mantz1509.00908}) and applied using an off-the-shelf Gibbs sampling environment \citep{Sereno1407.7869}.

A common and problematic feature of astronomical data that these specialized tools do not address is selection bias, resulting in truncation of the observed data set. This refers to the situation in which the data set available for analysis is not representative of the complete population that we wish to make inferences about, and furthermore that even the size of that complete population is not known.  Note that this scenario is distinct from that of ``censored'' data, in which a subset of measurements are unavailable even though the size of the complete data set is known and fixed. Modifications of classical nonparametric estimators have been developed to address truncation (\citealt{Efron1992ApJ...399..345E, Efron1999-2669997}). In the Bayesian framework, the solution is to incorporate the selection mechanism into a generative model for the data; this necessitates modeling the full population, including undetected (but potentially detectable) sources.
An unavoidable feature of inference on truncated data, which becomes explicit in the Bayesian formulation, is that we must have a model to describe the portion of the complete population that is not observed.

In astronomy, selection effects such as Eddington and Malmquist biases have been discussed at least since the eponymous works of \citet{Eddington1913MNRAS..73..359E} and \citet{Malmquist1922MeLuF.100....1M, Malmquist1925MeLuF.106....1M}. 
Their importance has been recognized for cosmological surveys, especially in the context of the abundance and scaling relations of clusters of galaxies (e.g., recently, \citealt{Pratt0809.3784, Vikhlinin0805.2207, Mantz0909.3098, Mantz0909.3099, Allen1103.4829}), and the distance-redshift relation of type Ia supernovae \citep{March1804.02474}. Similar selection effects are clearly also present in, for example, exoplanet surveys (e.g.\ \citealt{Youdin1105.1782, Gaidos1211.2279}), and have been discused in the context of quasar and gamma ray burst data sets \citep{Efron1994-2290845, Petrosian1504.01414}.
The discussion below thus has applicability in many areas.\footnote{Indeed, after this work was submitted and was in revision, \citet{Mandel1809.02063} wrote about a application of a similar framework to gravitational-wave astrophysics, with an emphasis on recovering the distribution of covariates.}

The purpose of this work is twofold. First, I hope to provide an understandable overview of how truncation can be incorporated into the likelihood function in general, as well as more concretely for a few specific (and simple) selection mechanisms. This will include some discussion of the special circumstances in which selection effects are ignorable, i.e.\ when the likelihood need not be modified.
The emphasis is on regression (that is, recovering the parameters of a linear relation), although the basic approach is more general.
Second, in simple cases where selection is not ignorable, constraints on a toy model obtained using the correct likelihood will be contrasted with those from methods analogous to the codes mentioned above, which do not account for selection. This is not to impugn those codes particularly, but to emphasize that failing to account for selection in an analysis has the potential to seriously compromise the results. In the conclusions, I will comment briefly on computational strategies for performing the complete analysis.
The code used to perform the fits to mock data in this work is available as an extension of the Python implementation of \lrgs{}.\footnote{\url{https://github.com/abmantz/lrgs}}

\section{A Concrete Scenario} \label{sec:scenario}

While many aspects of the model framework developed in the next section are general, it is helpful to have a specific problem in mind for illustration. Therefore, consider the closely related tasks of studying the cosmology and scaling relations of galaxy clusters. The key ingredients of the model, and the notation used in this work, are as follows.
\begin{itemize}
\item The population of clusters in the Universe is described theoretically by a mass function, i.e.\ their number density as a function of mass and redshift. The mass function is determined by cosmological parameters (e.g.\ \citealt{Press1974ApJ...187..425P}), which will be collectively denoted $\Omega$. Mass and redshift can be thought of as the covariates of a regression (below), and so are denoted $x$. Thus, the mass function, apart from a normalization, can be thought of as the a priori probability for a cluster to have a given mass and redshift, $p(x|\Omega)$. The normalization can be parametrized by $N$, the size of the complete population of interest. The interpretation of $N$ will depend on exactly what range of $x$ is used to define the population under study; in practice, the only requirement is that all sources that could plausibly be detected are included in this definition.
\item A given cluster generates various observable signals such as the mass, temperature, X-ray luminosity and Sunyaev-Zel'dovich signal of the intracluster gas; the number and optical/IR luminosity of its galaxies; and the gravitational lensing shear induced on background galaxies by the cluster's mass. These depend on the cluster mass and may evolve, and so can be thought of as response variables of a regression, $y$. Note that $y$ refers the true value of an observable quantity, not to the observed value, which is subject to measurement error. The average scaling of $y$ with $x$ is generally modeled as a power law (that is, a line if $x$ and $y$ actually refer to the logarithm of mass, etc.). In addition to this average behavior, there is an intrinsic scatter in the values of $y$ for a given $x$. The scaling relations are thus described by a distribution $p(y|x,\theta)$, where $\theta$ parametrizes both the average scaling relation(s) and the intrinsic scatter.
\end{itemize}
Here I have implicitly assumed that the parameters represented by $\Omega$ and $\theta$ are distinct. This need not always be true, but it is reasonably common separation, in this case reflecting a distinction between cosmological and astrophysical models.
\begin{itemize}
\item Measured (or potentially measured) values of the properties of a cluster will be denoted $\hat{x}$ or $\hat{y}$; these are related to the true values by a sampling distribution, $p(\hat{x},\hat{y}|x,y)$. This notation includes the possibility of correlations in the measurement errors. For simplicity, I will assume that the sampling distribution as a function of $x$ and $y$ is known, so that no additional parameters need to appear explicitly. The measured data additionally include $\Ndet$, the number of clusters detected in the survey.
\end{itemize}
In the galaxy cluster case, $\hat{x}$ could include spectroscopic measurements of redshift. However, measurements of mass are less straightforward, and in general any measured proxy for the mass may have an intrinsic scatter which correlates at fixed true mass with one of the observables in $y$. In practice, it therefore makes sense to include such mass proxies (including mass estimated from gravitational lensing) as response variables, with theoretical priors constraining the attendant scaling relation parameters. In practice, the same set of measurements need not be available for all detected clusters, with the exception of the survey measurement(s) used to detect them to begin with.
\begin{itemize}
\item The probability for a cluster to be detected (that is, included in the data set) as a function of measured (or potentially measured) properties is denoted $\Pdet(\hat{x},\hat{y}|\phi)$. This may depend on additional parameters, $\phi$, such as a completeness or a flux limit.
\end{itemize}

In practice, the detection process is generally a deterministic function of the survey data, so $\Pdet$ can be written as a function of only $\hat{x}$, $\hat{y}$ and $\phi$ for an appropriate definition of $x$ and $y$. In fact, it is frequently possible to express $\Pdet$ as a step function. For example, consider the scenario in which detection requires a measured flux to exceed a position-dependent threshold (corresponding to non-uniform survey depth). With position on the sky included in $x$ and flux included in $y$, $\Pdet$ has the form of a step function, dependent on position and measured flux. However, there is no real benefit to expressing things this way, since a cluster's position on the sky is typically both well determined (effectively without error) and not otherwise of interest. Hence, one might instead define $\Pdet$ in terms of measured flux only, with $\Pdet$ proportional to the fraction of the survey footprint where the threshold for detection is less than a given value. Conversely, a metric often used to characterize cluster surveys is the detection probability as a function of (true) mass. However, writing $\Pdet$ this way requires a marginalization over the dependent variable(s) associated with the survey detection (luminosity in the above example) as well as the corresponding scaling relation parameters. Consequently, this is not the most natural way to express $\Pdet$ in the likelihood developed in the next section.

To summarize this scenario, $N$ and $\Omega$ determine the number density of clusters as a function of redshift and mass ($x$); one or more response variables ($y$) for each cluster follow from its redshift, mass and the scaling relation parameters, $\theta$; potentially measured values $\hat{x}$ or $\hat{y}$ follow from $x$, $y$ and the sampling distribution; and these measurements combined with the detection probability result in $\Ndet$ clusters, and their measured values, forming the available data set. This level of generality will be maintained in Section~\ref{sec:theory}. In Section~\ref{sec:examples}, we will specify the scenario even further in order to illustrate the impact of selection effects on a toy data set.

\section{Theory} \label{sec:theory}

\subsection{Likelihood}
 
At its most abstract level, the problem at hand is that of modeling the properties of some population of sources in the Universe, when a fraction of that population is systematically missing from our data set. Using the notation introduced in Section~\ref{sec:scenario}, our model thus divides the $N$ sources in the complete population into $\Ndet$ that are represented in the data and $\Nmis=N-\Ndet$ that are missing from the data. If we assume that sources are independent from one another both in their occurrence and detection, the likelihood of the data, marginalized over $N$, can be written
\begin{eqnarray} \label{eq:mostgeneralL}
  \like = \sum_{N=\Ndet}^\infty p(N) {N \choose \Ndet} \Ldet \fmis^{\Nmis},
\end{eqnarray}
with $p(N)$ the prior distribution for $N$.
Here $\fmis$ is the a priori probability that a given source is \emph{not} detected, 
\begin{eqnarray} \label{eq:fmis}
  \fmis = \int dx \,dy \,d\hat{x} \,d\hat{y} \,p(x|\Omega)\, p(y|x,\theta) \,p(\hat{x},\hat{y}|x,y) \left[1-\Pdet(\hat{x},\hat{y}|\phi)\right],
\end{eqnarray}
which appears once for each of the $\Nmis$ undetected sources.
This construction explicitly shows the integrations required to express a completeness function (or effective $\Pdet$) in terms of true properties $x$ and/or $y$.
Note also that this framework requires the sampling distribution, $p(\hat{x},\hat{y}|x,y)$, to be defined for a generic source, at least for those observables involved in detection.
That is, we need a generative model for the measurement errors involved in the survey detection process, not just error bars for the detected sources estimated from the data.
$\Ldet$ is the likelihood associated with detections,
\begin{equation} \label{eq:ldet}
  \Ldet = \prod_{i=1}^{\Ndet} \int dx_i \,dy_i \,p(x_i|\Omega) \,p(y_i|x_i,\theta) \,p(\hat{x}_i,\hat{y}_i|x_i,y_i) \, \Pdet(\hat{x_i},\hat{y_i}|\phi),
\end{equation}
where the factorization into a product relies on our assumption that the sources occur independently.
The integral in this expression differs from that in Equation~\ref{eq:fmis} both in the substitution $1-\Pdet \rightarrow \Pdet$ and in that $\hat{x}_i$ and $\hat{y}_i$ are fixed by observation rather than being marginalized over. In the common case where $\phi$ is constant, $\Pdet(\hat{x_i},\hat{y_i}|\phi)$ is also a constant for all observed sources, and this factor is equivalent to the simplified likelihood (applicable in the absence of selection effects),
\begin{equation} \label{eq:lsimp}
  \Lsimp = \prod_{i=1}^{\Ndet} \int dx_i \,dy_i \,p(x_i|\Omega) \,p(y_i|x_i,\theta) \,p(\hat{x}_i,\hat{y}_i|x_i,y_i).
\end{equation}
The combinatoric factor in Equation~\ref{eq:mostgeneralL}, ${N \choose \Ndet}=N!/\Ndet!\Nmis!$, appears because the sources are a priori exchangeable.

\subsection{Prior Distributions for $N$} \label{sec:Npriors}

Since the complete population size, $N$, has been introduced, we will need to assign it a prior distribution. For example, \citet{Gelman2004BayesianDataAnalysis} note that if $p(N) \propto N^{-1}$ (uniform in $\log N$), the sum in Equation~\ref{eq:mostgeneralL} can be done analytically, yielding
\begin{eqnarray} \label{eq:Ninv_prior}
  \like = \Ldet \fdet^{-\Ndet},
\end{eqnarray}
with 
\begin{eqnarray} \label{eq:fdet}
  \fdet &=& 1-\fmis \\
  &=& \int dx \,dy \,d\hat{x} \,d\hat{y} \,p(x|\Omega) \,p(y|x,\theta) \,p(\hat{x},\hat{y}|x,y) \,\Pdet(\hat{x},\hat{y}|\phi) \nonumber
\end{eqnarray}
being the a priori probability for a source to be detected. Note that $\fdet$ has no dependence on the measured properties of the detected sources, despite the fact that it appears to the power $-\Ndet$.

While this identity makes the $N^{-1}$ prior convenient, a Poisson distribution (dependent on a mean hyperparameter, $\EN$) is more appropriate in most astronomical scenarios. This is consistent with our earlier assumption of independently occurring sources. 
With such a prior, Equation~\ref{eq:mostgeneralL} becomes
\begin{eqnarray} \label{eq:full_like}
  \like &=& \sum_{N=\Ndet}^\infty \frac{ e^{-\EN} \EN^N}{N!} \frac{N!}{\Ndet! \Nmis!} \fmis^{\Nmis} \Ldet, \\
        &=& \frac{e^{-\ENdet} \EN^{\Ndet}}{\Ndet!} \,\Ldet \sum_{\Nmis=0}^\infty \frac{e^{-\ENmis} \ENmis^{\Nmis}}{\Nmis!}, \nonumber\\
  &\propto& e^{-\fdet\EN} \EN^{\Ndet} \Ldet, \nonumber
\end{eqnarray}
where the identities $\fdet=\ENdet/\EN$ and $\fmis=\ENmis/\EN$ have been used, and where the last line discards a constant factor of $1/\Ndet!$. The same expression can be derived (perhaps more intuitively) without the need to explicitly model and marginalize over $N$ by considering the Poisson likelihood for sources in bins of the $\hat{x}$ and $\hat{y}$ observables, and taking the limit of infinitesimally small bins (see \citealt{Mantz0909.3098}).

In practice, $\EN$ may depend on further parameters of the astrophysical model (e.g.\ cosmological parameters in the galaxy cluster scenario of Section~\ref{sec:scenario}). However, for those cases where we lack a physically motivated prior for $\EN$, it may be convenient to assign a gamma distribution prior, as this makes the marginalization over both $N$ and $\EN$ analytic.
To see this, note that Equation~\ref{eq:full_like} has the form of a gamma distribution for $\EN$, with shape $\Ndet+1$ and rate $\fdet$. If we take a gamma prior on $\EN$ with shape $\alpha_o$ and rate $\beta_0$, then
\begin{eqnarray} \label{eq:like_w_gamma}
  p\left(\EN\right) \like &\propto& \mathrm{Gamma}\left(\EN \left| \Ndet+\alpha_0, \, \fdet+\beta_0\right.\right) \left(\fdet+\beta_0\right)^{-\left(\Ndet+\alpha_0\right)} \Ldet; \\
  \int_0^\infty d\EN \, p\left(\EN\right) \like &\propto& \left(\fdet+\beta_0\right)^{-\left(\Ndet+\alpha_0\right)} \Ldet, \nonumber
\end{eqnarray}
discarding constant factors that depend only on $\Ndet$, $\alpha_0$ and $\beta_0$. In the second line above, the gamma density function has integrated to unity, provided that $\alpha_0>-\Ndet$ and $\beta_0>-\fdet$.

Though not infinitely flexible, the gamma distribution provides a range of potentially useful priors. Taking $\beta_0 \rightarrow 0$, it describes power-law priors of the form $\EN^{\alpha_0-1}$. For both $\beta_0 \rightarrow 0$ and $\alpha_0\rightarrow0$ ($\EN^{-1}$, or uniform in $\log \EN$), we intuitively recover Equation~\ref{eq:Ninv_prior}, while $\beta_0\rightarrow0$ and $\alpha_0=1/2$ is the Jeffreys prior for $\EN$, $p\left(\EN\right) \propto \EN^{-1/2}$. When $\EN$ is expected to be large, approximately Gaussian priors can be accommodated by setting $\alpha_0=\mu^2/\sigma^2$ and $\beta_0=\mu/\sigma^2$, where $\mu$ and $\sigma$ are the desired mean and standard deviation. If desired, a posteriori samples of $\EN$ can be generated from the gamma distribution in Equation~\ref{eq:like_w_gamma}. A posteriori samples of $N$ could then be generated from a Poisson distribution; alternatively, if $\EN$ is not of interest, samples of $N$ can be drawn directly from its marginalized posterior distribution,
\begin{eqnarray}
  p(N) = \int d\EN \,p\left(\EN\right) p\left(N\left|\EN\right.\right) = \frac{\Gamma(N + \alpha_0)}{N! \, \Gamma(\alpha_0+1)} (\beta_0+1)^{-N} \left(\frac{\beta_0}{\beta_0+1}\right)^{\alpha_0},
\end{eqnarray}
which is the negative binomial distribution with parameters $\alpha_0$ and $(\beta_0+1)^{-1}$.

\subsection{Ignorability}

One of the central questions for this work is under what circumstances selection effects due to truncation require us to use Equation~\ref{eq:full_like}, rather than one of the simpler likelihoods $\Ldet$ or $\Lsimp$. The latter is possible when the posterior for the parameters of interest can be written strictly in terms of the observed data. \citet{Gelman2004BayesianDataAnalysis} refer to selection effects as {\it ignorable} in this case, and discuss the necessary conditions. To summarize, selection is ignorable if the following statements are both true:
\begin{enumerate}
\item The prior distribution for $\phi$ is independent of the prior distribution for all other parameters.
\item Selection does not depend on unobserved (or potentially unobserved) data.
\end{enumerate}
The first condition we can assume without losing too much generality, but the second is generically violated in truncation problems. Our default expectation in these circumstances should thus be that the formalism above is necessary. We will see below that, in very special circumstances, selection effects are ignorable for the purposes of constraining the parameters of the regression, $\theta$, though not necessarily $\Omega$ (assuming that the two are indeed separable). The extreme, and intuitive, example of this occurs when data are missing completely at random with respect to the measurements and parameters of interest; in that case, $\Ldet$ is naturally a sufficient likelihood.

Another way to put this is that using the likelihood $\Ldet$ alone is not the same as ``not using information from the number of detections'' -- that would be most closely equivalent to marginalizing $N$ over an uninformative prior, as outlined above.

\section{Simple Examples} \label{sec:examples}

\subsection{Toy Data, Models, and Methods} \label{sec:toy}

To illustrate how this works in a more concrete way, we can consider a simplified version of the galaxy cluster survey case outlined in Section~\ref{sec:scenario}. Specifically, let $x$ represent the log-mass only (neglecting redshift), and take
\begin{equation} \label{eq:pofx}
  p(x|\lambda) = \lambda e^{-\lambda x},
\end{equation}
with $\lambda=2$. This is an approximately appropriate distribution for the log-masses of galaxy clusters (e.g.\ \citealt{Evrard1403.1456}), apart from the unphysical restriction $x\geq0$. We will consider two response variables, $y=(y_1,y_2)$, with power-law slopes and an Gaussian intrinsic scatter covariance roughly appropriate for the log X-ray luminosity and log temperature of the intracluster gas, respectively \citep{Allen1103.4829, Giodini1305.3286};
\begin{eqnarray} \label{eq:scalingdist}
  p\left[\left.\left(\begin{array}{c}y_1\\y_2\end{array}\right)\right|x\right] &=& \normal\left[ \left.\left(\begin{array}{c}y_1\\y_2\end{array}\right)\right| \left(\begin{array}{c}\alpha_1\\\alpha_2\end{array}\right) + \left(\begin{array}{c}\beta_1\\\beta_2\end{array}\right) x, \,\left(\begin{array}{cc}\sigma^2_1 & \rho\sigma_1\sigma_2\\ \rho\sigma_1\sigma_2 & \sigma^2_2\end{array}\right) \right],
\end{eqnarray}
where $\normal$ denotes the multivariate normal density function for a given mean and covariance matrix. In particular, the marginal intrinsic scatter in $y_1$ at fixed $x$ is relatively large, and its average scaling is relatively steep, compared with the scatter and power-law slope of $y_2$, and the two scatters are moderately correlated. Measurement errors are assumed to be Gaussian, uncorrelated and identical for all sources, again with typical magnitudes, corresponding to respectably high signal-to-noise data;
\begin{eqnarray} \label{eq:measerr}
  p\left[\left.\left(\begin{array}{c}\hat{x}\\\hat{y}_1\\\hat{y}_2\end{array}\right)\right|\left(\begin{array}{c}x\\y_1\\y_2\end{array}\right)\right] &=& \normal\left[ \left. \left(\begin{array}{c}\hat{x}\\\hat{y}_1\\\hat{y}_2\end{array}\right)\right|\left(\begin{array}{c}x\\y_1\\y_2\end{array}\right), \,\left(\begin{array}{ccc}s^2_x & 0 & 0\\ 0 & s^2_{y_1} & 0\\ 0 & 0 & s^2_{y_2}\end{array}\right) \right].
\end{eqnarray}
Specifically, measurement errors were $s_x=0.2$ (roughly the intrinsic scatter due to correlated structure in cluster mass estimates from weak gravitational lensing; \citealt{Becker1011.1681}),\footnote{The assignment of a simple measurement error for mass contravenes the advice in Section~\ref{sec:scenario}, but is adopted for simplicity here.} $s_{y_1}=0.05$ and $s_{y_2}=0.1$. A complete (before truncation) mock data set of $10^4$ clusters was generated using these parameters, which are summarized in Table~\ref{tab:sim}. To make explicit the link to the notation of Sections~\ref{sec:scenario}--\ref{sec:theory}, we have $\Omega=\{\lambda\}$ and $\theta=\{\alpha_1,\alpha_2,\beta_1,\beta_2,\sigma_1,\sigma_2,\rho\}$.

\begin{table}
  \centering
  \caption[]{Model parameters used to generate the mock data set in Section~\ref{sec:examples}. See Equations~\ref{eq:pofx}--\ref{eq:measerr}.}
  \vspace{1ex}
  \begin{tabular}{cc@{\hspace{3em}}cc}
   Parameter & Value & Parameter & Value\\
    \hline
    $N$ & $10^4$ &      $\sigma_1$ & 0.4\\
    $\lambda$ & 2 &      $\sigma_2$ & 0.15\\
    $\alpha_1$ & 0 &     $\rho$ & 0.5\\
    $\alpha_2$ & 0 &     $s_x$ & 0.2\\
    $\beta_1$ & 1 &      $s_{y_1}$ & 0.05\\
    $\beta_2$ & 0.7 &    $s_{y_2}$ & 0.1\\
    \hline
  \end{tabular}
  \label{tab:sim}
\end{table}

The following subsections will apply a simple selection on either $\hat{y}_1$ or $\hat{x}$ and contrast constraints obtained using the complete likelihood of Equations~\ref{eq:full_like}--\ref{eq:like_w_gamma} with those obtained using only $\Ldet$ (equivalently, $\Lsimp$; Equations~\ref{eq:ldet}--\ref{eq:lsimp}). The constraints from $\Ldet$ were computed using a Python-language version of the \lrgs{} code that was straightforwardly extended to use the exponential form of $p(x)$ in Equation~\ref{eq:pofx} rather than the usual Gaussian mixture. Constraints from the full likelihood were found by alternating \lrgs{} conjugate-Gibbs sampling of the parameters that do not appear in $\fdet$ ($x_i$ and $y_i$) with Metropolis sampling (via the {\sc lmc} code\footnote{\url{https://github.com/abmantz/lmc}}) of the remaining parameters ($\lambda$, $\alpha_i$, $\beta_i$, $\sigma_i$ and $\rho$), a strategy implemented as a submodule of \lrgs{}. I will therefore refer to the two methods as \lrgs{} and \trunc{}, respectively.

Identical priors were applied to the parameters common to the two methods, specifically uniform priors for $\alpha_i$, $\beta_i$ and $\rho$; the Jeffreys prior for $\sigma_i^2$, $p(\sigma_i^2)\propto\sigma_i^{-2}$; and a Gaussian prior for $\lambda$, with mean 2 and standard deviation 0.05. For the \trunc{} method, I took an uninformative Gamma prior on $\EN$, with $\alpha_0=1/2$ and $\beta_0=0$, and followed the procedure in Section~\ref{sec:Npriors} to marginalize over $\EN$ analytically and generate samples a posteriori. These choices for $\lambda$ and $\EN$ priors mirror the typical situation in the analysis of galaxy cluster surveys, where we have prior information on the shape of the mass function, but wish to either fit for or marginalize over its normalization.

\subsection{Selection on the Survey Response Variable} \label{sec:dependent}

\begin{figure*}
  \centering
  \includegraphics{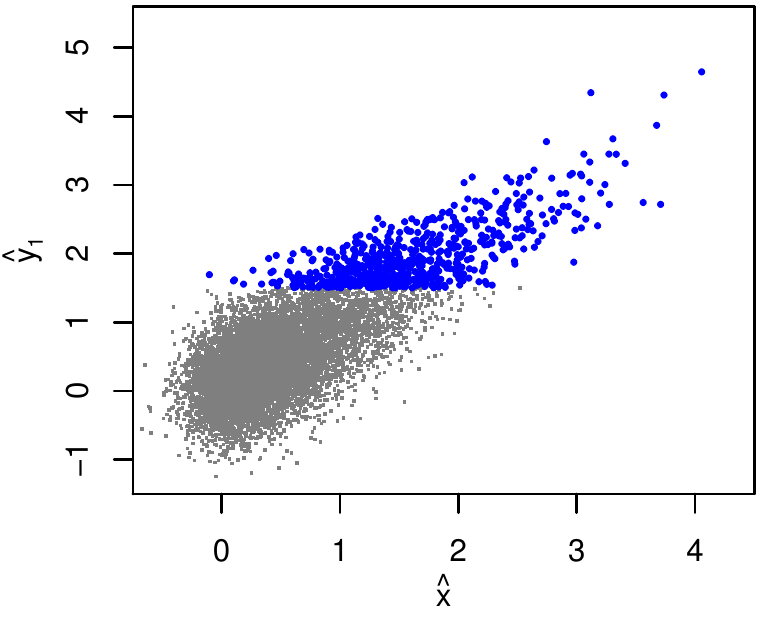}
  \hspace{2mm}
  \includegraphics{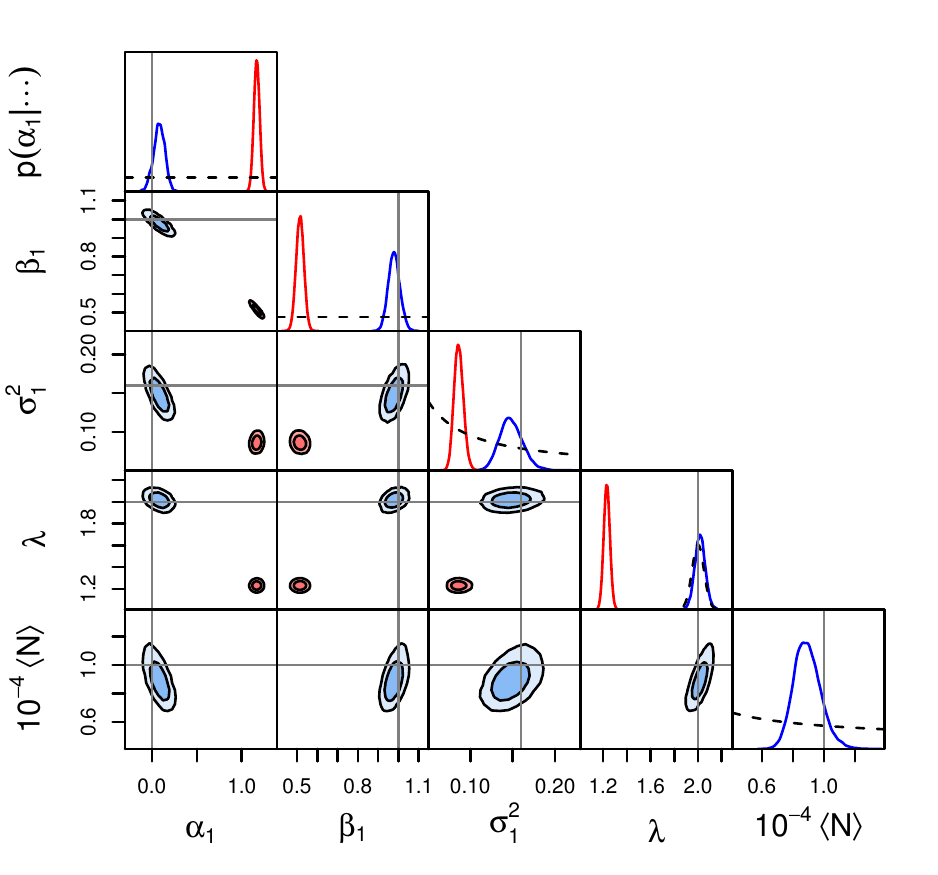}
  \caption{
    Mock data and constraints for the example illustrating truncation on $\hat{y}_1$ (Section~\ref{sec:dependent}). Left: Blue points show the observed data set, while gray points are unobserved.
    Right: Parameter constraints from the \lrgs{} (red) and \trunc{} (blue) methods (see Section~\ref{sec:toy}). Panels on the diagonal show marginalized posteriors for each parameter (solid curves), along with the prior (dashed curves; for improper priors these are arbitrarily normalized). Off-diagonal panels show the 68.3 and 95.4 per cent confidence regions obtained on each pair of parameters from each method. The input parameter values used to generate the mock data are indicated by solid, gray lines in each panel. Note that the \lrgs{} model does not include the $\EN$ parameter and hence does not produce constraints on it.
  } \label{fig:truncy}
\end{figure*}

When the scaling relation of interest is for the dependent variable on which selection is based, it is clear that the requirements for selection to be ignorable are not met (Section~\ref{sec:theory}). Consider the simple detection requirement $\hat{y}_1 > \ylim$, i.e.\ $\Pdet(\hat{y}_1|\ylim) = \Theta(\hat{y}_i-\ylim)$, with $\Theta$ the unit step function. Figure~\ref{fig:truncy} illustrates this selection on the mock data set with $\ylim=1.5$, for which 658 points are ``detected'' in this particular realization. This is a sufficiently large data set that the systematic error introduced by using an incorrect likelihood is significant compared with the width of the posterior. In the first case, consider a fit only involving $x$ and $y_1$, ignoring any information about $y_2$ (but see Section~\ref{sec:correlated}).

For the particular scenario described above, we have
\begin{equation} \label{eq:fdet_truncy}
  \fdet = \int dx \, \lambda e^{-\lambda x} \left[1 - \Phi\left( \frac{\ylim-\alpha_1-\beta_1 x}{\sqrt{\sigma_1^2 + s_{y_1}^2}} \right)\right],
\end{equation}
where $\Phi$ is the standard normal cumulative distribution function. More generally, selection on $\hat{y}_1$ implies that $\fdet$ will depend explicitly on the parameters governing the marginal scaling relations of $y_1$; hence, the correct posterior for these parameters cannot be recovered if terms in the likelihood involving $\fdet$ are neglected. Schemes that employ only ``bias corrections'' of the sampling distribution, by setting $p(\hat{y}_1|y_1)$ to zero below $\hat{y}_1=\ylim$ and renormalizing it (e.g.\ \citealt{Vikhlinin0805.2207, Sereno1407.7869}), do not address this feature. Note that both the methods considered here already implicitly include this information, since $P(\hat{y_1}<\ylim)=0$ for all detected sources.

The constraints obtained from \lrgs{} and \trunc{} are respectively shown as red and blue contours in the right panel of Figure~\ref{fig:truncy}; evidently, the former disagree with the input parameter values at high significance.

How can we intuitively understand this? First, it's worth noting that prior information about the form of $p(x)$ in has nothing to do with the bias in the constraints from \lrgs{}. In fact, fixing $\lambda$ (which \lrgs{} gets spectacularly wrong, in spite of the informative prior) to the true value does not significantly change the constraints on the scaling parameters. More formally, if we take the limit of zero measurement errors on $\hat{x}$, the likelihood $\Lsimp$ (or $\Ldet$) provides no mechanism to produce covariance between $\lambda$ and the scaling parameters. We should therefore expect the bias produced by neglecting terms with $\fdet$ in the likelihood to persist, even with perfect prior information on $p(x)$, despite the fact that the true $p(x)$ clearly implies that a substantial number of objects must be missing at $x\ltsim2$.

The reason that $\Lsimp$ cannot recover the input scaling relation, even with accurate prior information about $p(x)$, is that it fails to capture the \emph{systematic} way in which sources with small $x$ are missing, specifically the dependence on their value of $\hat{y}_1$. The practice of truncating the sampling distribution for $\hat{y}_1$ at $\ylim$ is of no help here; while it may prevent the cluster of observed points just above $\ylim$ at small $x$ from significantly penalizing models near the truth, models with shallower slopes that pass closer to these points will still be preferred. This is exactly what we see in the constraints from \lrgs{}.

In contrast, the \trunc{} method, when provided with the same prior information, recovers the true parameter values. In this case, the highly biased points detected at small $x$ do carry significant information. Their particular values of $\hat{y}_1$ are not very informative for models near the truth -- $\ylim$ is so far from the mean scaling relation that detected points \emph{must} lie just above the threshold. But the \emph{number} of sources that exceed $\ylim$, combined with knowledge of $\expectation{dN/dx}$, constrains the scaling relation and its scatter. In this example, there are 100 detections with $x<1$, implying (for $\EN=10^4$) that $\ylim$ exceeds the mean relation by $\sim 2$--$3\,\sigma_1$ in this regime. This dependence of the interpretation of the data on $\EN$ is illustrated by the degeneracies between $\EN$ and the other parameters of interest in Figure~\ref{fig:truncy}.

\subsection{Selection on the Covariate} \label{sec:independent}

At the other extreme, consider selection on $\hat{x}$ instead of $\hat{y}_1$, $\Pdet(\hat{x}|\xlim) = \Theta(\hat{x}-\xlim)$, again ignoring $y_2$ for the moment. In this case, $\fdet$ has an analogous form to Equation~\ref{eq:fdet_truncy},
\begin{equation} \label{eq:fdet_truncx}
  \fdet = \int dx \, \lambda e^{-\lambda x} \left[1 - \Phi\left( \frac{\xlim-x}{s_x} \right)\right].
\end{equation}
Intuitively, the detected fraction is now independent of the scaling relation parameters, $\theta$. It follows that:
\begin{itemize}
\item If our model for $\expectation{dN/dx}$ is fixed a priori, then $e^{-\fdet\EN} \EN^{\Ndet}$ is a constant and selection effects are ignorable. This holds regardless of whether there are non-zero measurement errors. If $\EN$ is a free parameter, selection is still ignorable \emph{for inferences about $\theta$} because the likelihood factors into one part that depends on $\EN$ and another that depends on $\theta$, with no free parameters appearing in both.
\item If measurement errors on $\hat{x}$ are zero (the latent parameters $x$ are effectively fixed), then selection is ignorable for inferences about $\theta$ (only). This is because the observed data are always complete and therefore unbiased for every $x$ that is represented in the data set.
\end{itemize}
The above are special cases, however. In general, when there are non-zero measurement errors on $\hat{x}$ and the model for $p(x)$ is not fixed, selection must be accounted for. Note that even these exceptions depend on the parameters governing $p(x)$ and the scaling relation ($\Omega$ and $\theta$) being distinct, which was an assumption in this application (Section~\ref{sec:scenario}), but is not true of  all possible applications.

\begin{figure*}
  \centering
  \includegraphics{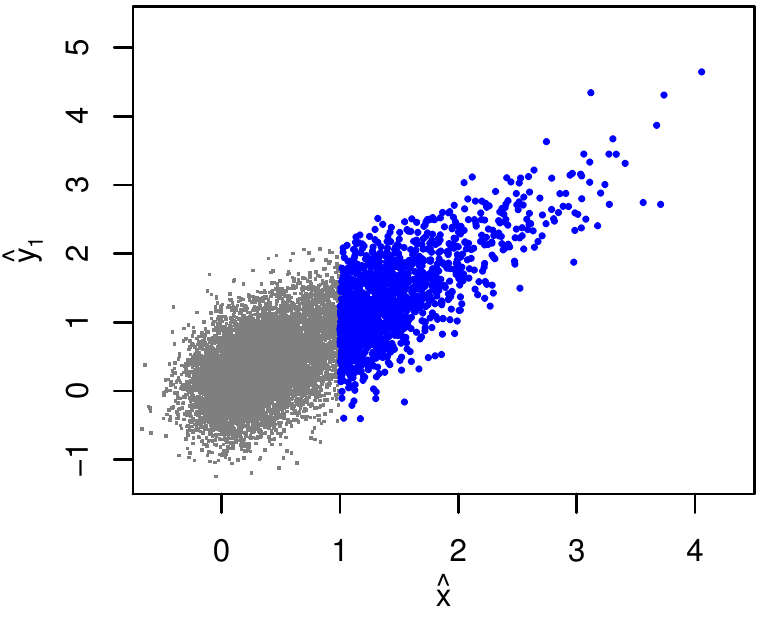}
  \hspace{2mm}
  \includegraphics{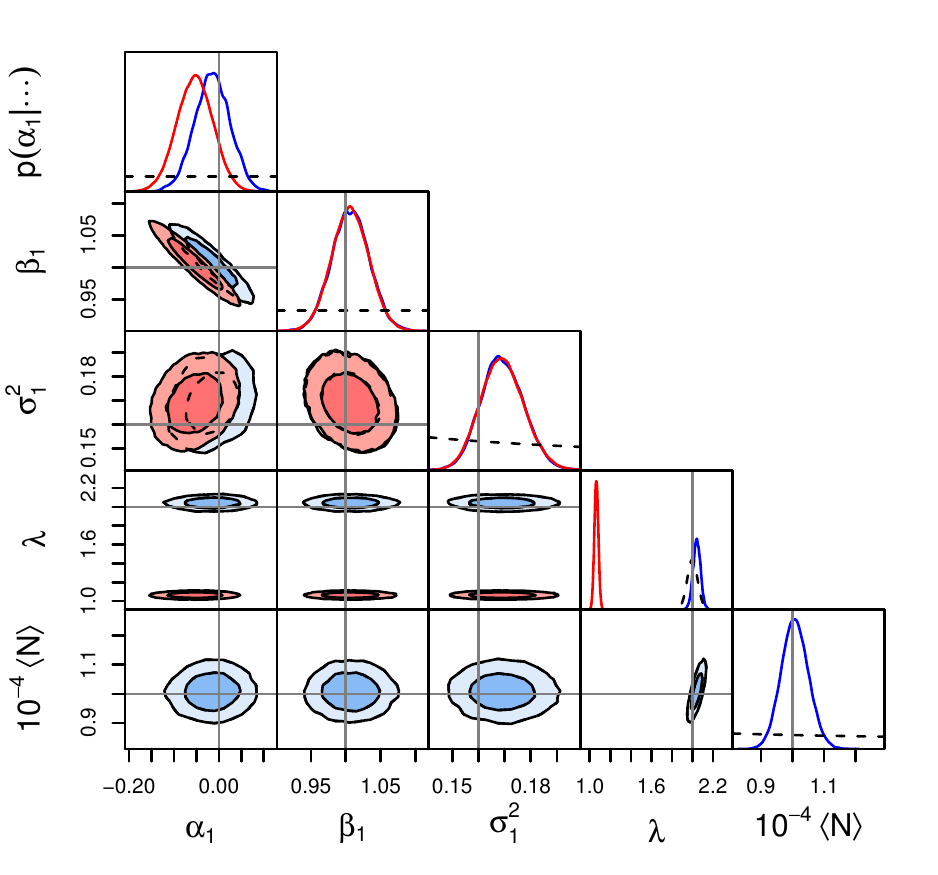}
  \caption{
    As Figure~\ref{fig:truncy}, but for the example illustrating truncation on $\hat{x}$ (Section~\ref{sec:independent}).
  } \label{fig:truncx}
\end{figure*}

The mock data selection for $\xlim=1$ ($\Ndet=1419$) and the resulting constraints appear in Figure~\ref{fig:truncx}. As one might guess from the discussion above, the constraints from \lrgs{} are less biased than before, although the joint posterior for $\alpha_1$ and $\beta_1$ is still inconsistent with the input parameters at high significance. Again, fixing the value of $\lambda$ would not eliminate the bias on the other parameters (see above).

\subsection{Selection on a Correlated Response Variable} \label{sec:correlated}

Next, consider the case where we are interested in the scaling of $y_2$ with $x$ when the data set is selected on a different response variable, $\hat{y}_1$. The key question here is whether $\hat{y}_1$ and $\hat{y}_2$ are correlated at fixed $x$, either due to correlation of their measurement errors or due to an intrinsic covariance of $y_1$ and $y_2$ at fixed $x$. If no such correlation is possible, then selection on $\hat{y}_1$ is equivalent to a (possibly noisy) selection on $x$, and the comments in Section~\ref{sec:independent} apply.

For illustration, Equation~\ref{eq:scalingdist} can be rewritten as
\begin{equation} \label{eq:mvscaling}
  p\left[\left.\left(\begin{array}{c}y_1\\y_2\end{array}\right)\right|x\right] = \normal\left(y_1\left|x,\sigma_1^2\right.\right) \normal\left[y_2\left| \alpha_2+\beta_2x+\rho\frac{\sigma_2}{\sigma_1}(y_1-\alpha_1-\beta_1 x), \,\sigma_2^2(1-\rho^2) \right.\right].
\end{equation}
This factorization demonstrates how our interpretation of $y_2$ may depend on information about $y_1$, such as satisfaction of a selection criterion, for $\rho\neq0$.
Specifically, the difference between $y_1$ and its mean value predicted by the scaling relation, $\alpha_1+\beta_1 x$, impacts our interpretation of the analogous displacement of $y_2$ from its mean scaling law.
Thus, for a positive correlation, at low masses ($x$) we expect a selection on luminosity ($y_1$) to bias the observed data high in both luminosity and temperature ($y_2$).
The detected fraction for selection on $\hat{y}_1$ is given by Equation~\ref{eq:fdet_truncy}.

As in Section~\ref{sec:independent}, we can see that selection effects are ignorable for inference of $\alpha_2$, $\beta_2$ and $\sigma_2$ only in very special circumstances, namely if
\begin{enumerate}
\item the marginal scaling relation for $y_1$ (i.e.\ the values of $\alpha_1$, $\beta_1$ and $\sigma_1$) and the intrinsic correlation coefficient, $\rho$, are fixed a priori; \emph{and}
\item the model for $p(x)$ is fixed or the measurement errors for $\hat{x}$ are zero.
\end{enumerate}
Note that the second condition is identical to the requirement for selection on $\hat{x}$ to be ignorable. The requirements of the first condition above are exactly those that make selection on $\hat{y}_1$ equivalent to a noisy selection on $\hat{x}$, with the nature of that stochasticity fully understood.

\begin{figure*}
  \centering
  \includegraphics{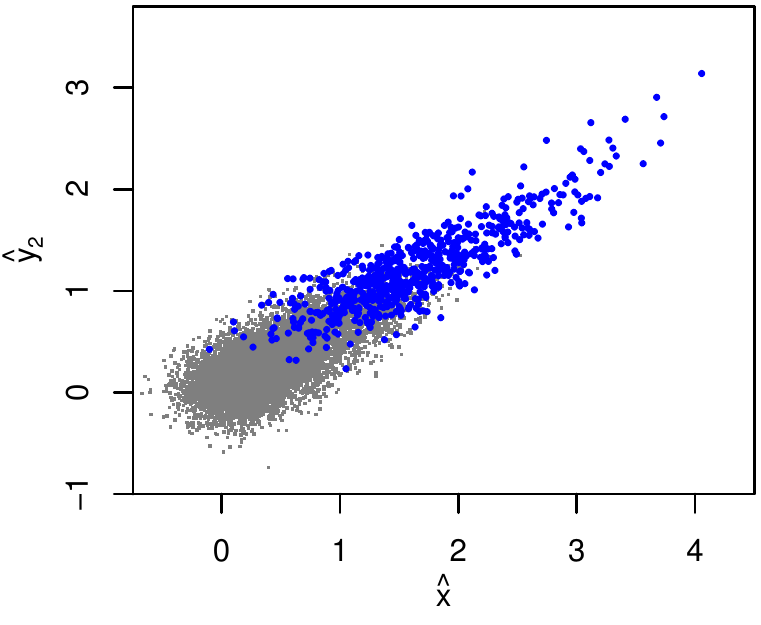}
  \caption{
    As Figure~\ref{fig:truncy} (left panel), but showing the $\hat{x}$ and $\hat{y}_2$ values when selection is on $\hat{y}_1$ (Section~\ref{sec:correlated}) The intrinsic scatter in $y_1$ and $y_2$ at fixed $x$ is moderately correlated.
  } \label{fig:truncy2_xy}
\end{figure*}

Using the same selection as in Section~\ref{sec:dependent} ($\ylim>1.5$), Figure~\ref{fig:truncy2_xy} shows the complete and observed mock data set in terms of $\hat{x}$ and $\hat{y}_2$. Due to the modest intrinsic correlation ($\rho=0.5$) and relatively smaller marginal scatter $\sigma_2$ compared with $\sigma_1$, selection effects on the observed data are less visually dramatic than in Figures~\ref{fig:truncy}--\ref{fig:truncx}. Nevertheless, we will see below that neglecting to model the truncation results in biased inferences.

\begin{figure*}
  \centering
  \includegraphics{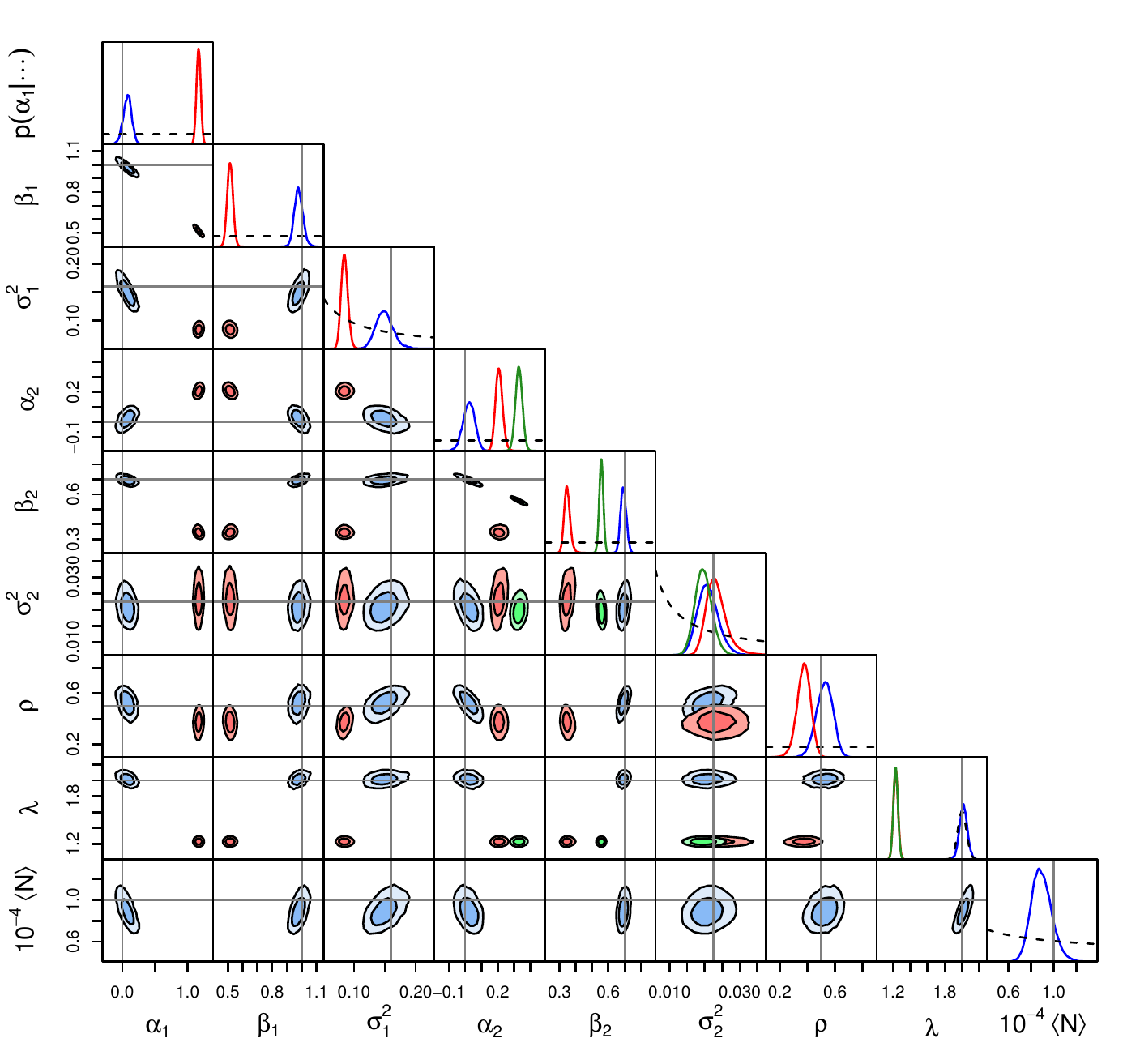}
  \caption{
    As Figure~\ref{fig:truncy} (right panel), but for the example illustrating a simultaneous fit for the scaling relations of $y_1$ and $y_2$ with selection on $\hat{y}_1$ (Section~\ref{sec:correlated}). Green curves and contours correspond to an \lrgs{} fit in which $y_1$ is neglected completely.
  } \label{fig:truncy2}
\end{figure*}

Figure~\ref{fig:truncy2} compares constraints from \lrgs{} and \trunc{} in the usual way, where both codes are now fitting joint scaling relations and scatter for $y_1$ and $y_2$ as a function of $x$. In addition, constraints are shown from an \lrgs{} analysis where $y_1$ is disregarded completely, i.e.\ simply fitting $y_2$ against $x$ without accounting for selection effects. In this particular case, both \lrgs{} analyses are consistent with the input value of $\sigma_2$, as one might guess by inspection of Figure~\ref{fig:truncy2_xy}, but produce biased constraints (to differing degrees) on $\alpha_2$ and $\beta_2$.

Note that a non-zero correlation in the measurement errors of $y_1$ and $y_2$ would play essentially the same role as the intrinsic correlation, $\rho$, in the discussion above.

\section{Discussion and Conclusions}

Although the examples explored above are far from exhaustive, hopefully it's clear that selection effects have the potential to dramatically bias the results of otherwise straightforward model fitting if not taken into account. Exactly how important this systematic effect is compared with the statistical uncertainties is not a simple question to answer in general, as it will depend not just on the number of observed data points and their error bars, but also on the selection mechanism and the true, underlying model. One could always straightforwardly test whether simple fitting methods are able to recover the correct parameter values by running them on mock data appropriate for a given situation, along the lines of the examples above. A better option, whenever feasible, would be to properly include selection in the model being fit.

An unfortunate feature of models that account for selection is that they lack the full conjugacy that allows all of the parameters in models like those used by \lrgs{} and {\sc linmix\_err} to be efficiently Gibbs sampled, even for simple selection mechanisms like those considered here. Specifically, conjugacy will generally be lost for any parameters appearing in $\fdet$. It is, however, still possible to efficiently Gibbs sample the remaining parameters, in particular $x_i$ and $y_i$, under the assumptions made by these codes, namely Gaussian (or similarly convenient) forms of the measurement errors, intrinsic scatter, and $p(x)$. Since $x_i$ and $y_i$ normally account for the great majority of the free parameters in such models, mixing conjugate Gibbs sampling of with some other method of sampling the remaining parameters, as outlined in Section~\ref{sec:toy}, is a viable strategy for these cases (though I by no means claim it to be the most efficient strategy). Note that this strategy is not without its pitfalls; in particular, when using mixture models, the potentially large number of parameters and the exchangeability of the mixture components can make sampling challenging (this is a generic feature, not specific to truncation problems). In addition, it's potentially helpful that $\EN$ can be marginalized analytically for a wide range of approximately power-law and Gaussian priors.

\begin{figure*}
  \centering
  \includegraphics{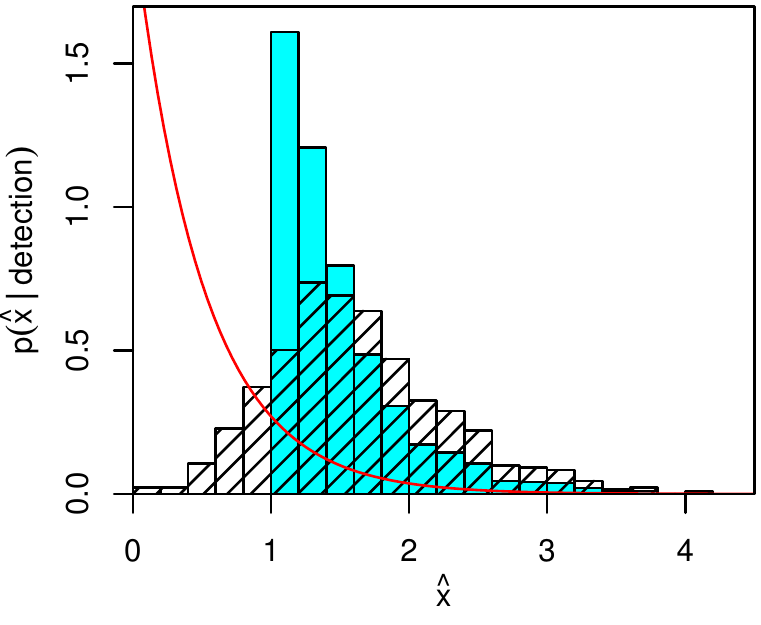}
  \caption{
    Distributions of $\hat{x}$ for the mock data sets selected by requiring $\hat{y}_1>1.5$ (hatched) or $\hat{x}>1$ (cyan shaded). The red curve shows the true $p(x)$ describing the complete data set.
  } \label{fig:px}
\end{figure*}

A basic and intuitive feature of truncation is that our interpretation of the data relies to some extent on a model for the population of sources that were not observed. There are two immediate consequences of this. Firstly, we can expect our results in general to be sensitive to prior information about $\expectation{dN/dx}$, including the form of $p(x)$. Thus, the Gaussian mixture models employed by some ``out of the box'' codes, while convenient and flexible, are no substitute for accurate modeling of $p(x)$. Inspection of the distributions of $\hat{x}$ selected from the mock data sets analyzed above (Figure~\ref{fig:px}) makes clear that no amount of flexible but uninformed modeling of the observed $\hat{x}$ data is likely to recover or even be consistent with the underlying, non-Gaussian form of $p(x)$. While we may not need precise knowledge of the true $p(x)$ a priori to obtain correct results in this example, we would likely at least need the prior that $p(x)$ is monotonically decreasing. Secondly, the amount of data required for our results to be data-dominated rather than prior-dominated will generally be greater than in problems without truncation, and may not be particularly obvious. Analysis of mock data sets is probably the best way to get a handle on this.

Despite these complicating aspects, the general solution for fitting truncated data is relatively straightforward. This is encouraging, given that truncation is such a common feature of astrophysical data.

\section*{Acknowledgments}
This work was supported by the National Aeronautics and Space Administration under Grant No.\ NNX15AE12G issued through the ROSES 2014 Astrophysics Data Analysis Program. I thank Gus Evrard and Arya Farahi for interesting discussions, and the anonymous referee for very good suggestions.


\begin{thebibliography}{}
\makeatletter
\relax
\def\mn@urlcharsother{\let\do\@makeother \do\$\do\&\do\#\do\^\do\_\do\%\do\~}
\def\mn@doi{\begingroup\mn@urlcharsother \@ifnextchar [ {\mn@doi@}
  {\mn@doi@[]}}
\def\mn@doi@[#1]#2{\def\@tempa{#1}\ifx\@tempa\@empty \href
  {http://dx.doi.org/#2} {doi:#2}\else \href {http://dx.doi.org/#2} {#1}\fi
  \endgroup}
\def\mn@eprint#1#2{\mn@eprint@#1:#2::\@nil}
\def\mn@eprint@arXiv#1{\href {http://arxiv.org/abs/#1} {{\tt arXiv:#1}}}
\def\mn@eprint@dblp#1{\href {http://dblp.uni-trier.de/rec/bibtex/#1.xml}
  {dblp:#1}}
\def\mn@eprint@#1:#2:#3:#4\@nil{\def\@tempa {#1}\def\@tempb {#2}\def\@tempc
  {#3}\ifx \@tempc \@empty \let \@tempc \@tempb \let \@tempb \@tempa \fi \ifx
  \@tempb \@empty \def\@tempb {arXiv}\fi \@ifundefined
  {mn@eprint@\@tempb}{\@tempb:\@tempc}{\expandafter \expandafter \csname
  mn@eprint@\@tempb\endcsname \expandafter{\@tempc}}}

\bibitem[\protect\citeauthoryear{{Allen}, {Evrard}  \& {Mantz}}{{Allen}
  et~al.}{2011}]{Allen1103.4829}
{Allen} S.~W.,  {Evrard} A.~E.,   {Mantz} A.~B.,  2011, \mn@doi [\araa]
  {10.1146/annurev-astro-081710-102514}, \href
  {http://adsabs.harvard.edu/abs/2011ARA%26A..49..409A} {49, 409}

\bibitem[\protect\citeauthoryear{{Becker} \& {Kravtsov}}{{Becker} \&
  {Kravtsov}}{2011}]{Becker1011.1681}
{Becker} M.~R.,  {Kravtsov} A.~V.,  2011, \mn@doi [\apj]
  {10.1088/0004-637X/740/1/25}, \href
  {http://adsabs.harvard.edu/abs/2011ApJ...740...25B} {740, 25}

\bibitem[\protect\citeauthoryear{{Eddington}}{{Eddington}}{1913}]{Eddington1913MNRAS..73..359E}
{Eddington} A.~S.,  1913, \mn@doi [\mnras] {10.1093/mnras/73.5.359}, \href
  {http://adsabs.harvard.edu/abs/1913MNRAS..73..359E} {73, 359}

\bibitem[\protect\citeauthoryear{{Efron} \& {Petrosian}}{{Efron} \&
  {Petrosian}}{1992}]{Efron1992ApJ...399..345E}
{Efron} B.,  {Petrosian} V.,  1992, \mn@doi [\apj] {10.1086/171931}, \href
  {http://adsabs.harvard.edu/abs/1992ApJ...399..345E} {399, 345}

\bibitem[\protect\citeauthoryear{Efron \& Petrosian}{Efron \&
  Petrosian}{1994}]{Efron1994-2290845}
Efron B.,  Petrosian V.,  1994, \jasa, \href
  {http://www.jstor.org/stable/2290845} {89, 452}

\bibitem[\protect\citeauthoryear{Efron \& Petrosian}{Efron \&
  Petrosian}{1999}]{Efron1999-2669997}
Efron B.,  Petrosian V.,  1999, \jasa, \href
  {http://www.jstor.org/stable/2669997} {94, 824}

\bibitem[\protect\citeauthoryear{{Evrard}, {Arnault}, {Huterer}  \&
  {Farahi}}{{Evrard} et~al.}{2014}]{Evrard1403.1456}
{Evrard} A.~E.,  {Arnault} P.,  {Huterer} D.,   {Farahi} A.,  2014, \mn@doi
  [\mnras] {10.1093/mnras/stu784}, \href
  {http://adsabs.harvard.edu/abs/2014MNRAS.441.3562E} {441, 3562}

\bibitem[\protect\citeauthoryear{{Gaidos} \& {Mann}}{{Gaidos} \&
  {Mann}}{2013}]{Gaidos1211.2279}
{Gaidos} E.,  {Mann} A.~W.,  2013, \mn@doi [\apj] {10.1088/0004-637X/762/1/41},
  \href {http://adsabs.harvard.edu/abs/2013ApJ...762...41G} {762, 41}

\bibitem[\protect\citeauthoryear{{Gelman}, {Carlin}, {Stern}  \&
  {Rubin}}{{Gelman} et~al.}{2004}]{Gelman2004BayesianDataAnalysis}
{Gelman} A.,  {Carlin} J.~B.,  {Stern} H.~S.,   {Rubin} D.~B.,  2004, Bayesian
  Data Analysis.
Chapman \& Hall/CRC, \url {http://www.stat.columbia.edu/~gelman/book/}

\bibitem[\protect\citeauthoryear{{Giodini}, {Lovisari}, {Pointecouteau},
  {Ettori}, {Reiprich}  \& {Hoekstra}}{{Giodini}
  et~al.}{2013}]{Giodini1305.3286}
{Giodini} S.,  {Lovisari} L.,  {Pointecouteau} E.,  {Ettori} S.,  {Reiprich}
  T.~H.,   {Hoekstra} H.,  2013, \mn@doi [\ssr] {10.1007/s11214-013-9994-5},
  \href {http://adsabs.harvard.edu/abs/2013SSRv..177..247G} {177, 247}

\bibitem[\protect\citeauthoryear{{Kelly}}{{Kelly}}{2007}]{Kelly0705.2774}
{Kelly} B.~C.,  2007, \mn@doi [\apj] {10.1086/519947}, \href
  {http://adsabs.harvard.edu/abs/2007ApJ...665.1489K} {665, 1489}

\bibitem[\protect\citeauthoryear{{Malmquist}}{{Malmquist}}{1922}]{Malmquist1922MeLuF.100....1M}
{Malmquist} K.~G.,  1922, Meddelanden fran Lunds Astronomiska Observatorium
  Serie I, \href {http://adsabs.harvard.edu/abs/1922MeLuF.100....1M} {100, 1}

\bibitem[\protect\citeauthoryear{{Malmquist}}{{Malmquist}}{1925}]{Malmquist1925MeLuF.106....1M}
{Malmquist} K.~G.,  1925, Meddelanden fran Lunds Astronomiska Observatorium
  Serie I, \href {http://adsabs.harvard.edu/abs/1925MeLuF.106....1M} {106, 1}

\bibitem[\protect\citeauthoryear{{Mandel}, {Farr}  \& {Gair}}{{Mandel}
  et~al.}{2018}]{Mandel1809.02063}
{Mandel} I.,  {Farr} W.~M.,   {Gair} J.~R.,  2018, preprint, \href
  {http://adsabs.harvard.edu/abs/2018arXiv180902063M} {} (\mn@eprint {arXiv}
  {1809.02063})

\bibitem[\protect\citeauthoryear{{Mantz}}{{Mantz}}{2016}]{Mantz1509.00908}
{Mantz} A.~B.,  2016, \mn@doi [\mnras] {10.1093/mnras/stv3008}, \href
  {http://adsabs.harvard.edu/abs/2016MNRAS.457.1279M} {457, 1279}

\bibitem[\protect\citeauthoryear{{Mantz}, {Allen}, {Rapetti}  \&
  {Ebeling}}{{Mantz} et~al.}{2010a}]{Mantz0909.3098}
{Mantz} A.,  {Allen} S.~W.,  {Rapetti} D.,   {Ebeling} H.,  2010a, \mn@doi
  [MNRAS] {10.1111/j.1365-2966.2010.16992.x}, \href
  {http://adsabs.harvard.edu/abs/2010MNRAS.406.1759M} {406, 1759}

\bibitem[\protect\citeauthoryear{{Mantz}, {Allen}, {Ebeling}, {Rapetti}  \&
  {Drlica-Wagner}}{{Mantz} et~al.}{2010b}]{Mantz0909.3099}
{Mantz} A.,  {Allen} S.~W.,  {Ebeling} H.,  {Rapetti} D.,   {Drlica-Wagner} A.,
   2010b, \mn@doi [MNRAS] {10.1111/j.1365-2966.2010.16993.x}, \href
  {http://adsabs.harvard.edu/abs/2010MNRAS.406.1773M} {406, 1773}

\bibitem[\protect\citeauthoryear{{March}, {Wolf}, {Sako}, {D'Andrea}  \&
  {Brout}}{{March} et~al.}{2018}]{March1804.02474}
{March} M.,  {Wolf} R.,  {Sako} m.,  {D'Andrea} C.,   {Brout} D.,  2018,
  preprint, \href {http://adsabs.harvard.edu/abs/2018arXiv180402474M} {}
  (\mn@eprint {arXiv} {1804.02474})

\bibitem[\protect\citeauthoryear{{Petrosian}, {Kitanidis}  \&
  {Kocevski}}{{Petrosian} et~al.}{2015}]{Petrosian1504.01414}
{Petrosian} V.,  {Kitanidis} E.,   {Kocevski} D.,  2015, \mn@doi [\apj]
  {10.1088/0004-637X/806/1/44}, \href
  {http://adsabs.harvard.edu/abs/2015ApJ...806...44P} {806, 44}

\bibitem[\protect\citeauthoryear{{Pratt}, {Croston}, {Arnaud}  \&
  {B{\"o}hringer}}{{Pratt} et~al.}{2009}]{Pratt0809.3784}
{Pratt} G.~W.,  {Croston} J.~H.,  {Arnaud} M.,   {B{\"o}hringer} H.,  2009,
  \mn@doi [\aap] {10.1051/0004-6361/200810994}, \href
  {http://adsabs.harvard.edu/abs/2009A%26A...498..361P} {498, 361}

\bibitem[\protect\citeauthoryear{{Press} \& {Schechter}}{{Press} \&
  {Schechter}}{1974}]{Press1974ApJ...187..425P}
{Press} W.~H.,  {Schechter} P.,  1974, \mn@doi [\apj] {10.1086/152650}, \href
  {http://adsabs.harvard.edu/abs/1974ApJ...187..425P} {187, 425}

\bibitem[\protect\citeauthoryear{{Sereno}, {Ettori}  \& {Moscardini}}{{Sereno}
  et~al.}{2015}]{Sereno1407.7869}
{Sereno} M.,  {Ettori} S.,   {Moscardini} L.,  2015, \mn@doi [\mnras]
  {10.1093/mnras/stv809}, \href
  {http://adsabs.harvard.edu/abs/2015MNRAS.450.3649S} {450, 3649}

\bibitem[\protect\citeauthoryear{{Vikhlinin} et~al.,}{{Vikhlinin}
  et~al.}{2009}]{Vikhlinin0805.2207}
{Vikhlinin} A.,  et~al., 2009, \mn@doi [\apj] {10.1088/0004-637X/692/2/1033},
  \href {http://adsabs.harvard.edu/abs/2009ApJ...692.1033V} {692, 1033}

\bibitem[\protect\citeauthoryear{{Youdin}}{{Youdin}}{2011}]{Youdin1105.1782}
{Youdin} A.~N.,  2011, \mn@doi [\apj] {10.1088/0004-637X/742/1/38}, \href
  {http://adsabs.harvard.edu/abs/2011ApJ...742...38Y} {742, 38}

\makeatother
\end{thebibliography}
\def \jasa {J.\ Am.\ Stat.\ Assoc.} 

\end{document}